\documentclass[%
reprint,
 nofootinbib,
 amsmath,amssymb,
 aps,
prd,
 floatfix,
]{revtex4-1}

\usepackage{graphicx}
\usepackage{setspace}
\usepackage{color}
\usepackage{dcolumn}
\usepackage{bm}
\usepackage{subcaption}
\usepackage{hyperref}

\begin{document}


\title{``Rosenbluth'' separation of the $J/\psi $ near-threshold photoproduction -
	an access to the gluon Gravitational Form Factors at high $t$} 

\author{Lubomir~Pentchev}
\altaffiliation{Thomas Jefferson National Accelerator Facility, Newport News, Virginia 23606, USA}
\email{pentchev@jlab.org}
\author{Eugene~Chudakov}
\altaffiliation{Thomas Jefferson National Accelerator Facility, Newport News, Virginia 23606, USA}
\email{gen@jlab.org}




\begin{abstract}
We perform analysis of the near-threshold $J/\psi $ photoproduction data off the proton
based on two theoretical approaches, 
GPD \cite{Guo3} and holographic \cite{Zahed2},
that represent  the differential cross sections as powers of the skewness
parameter with coefficients that depend only on the momentum transfer $t$.
This allows to separate kinematically the corresponding coefficient functions 
in much the same way
as this is done for the electric and magnetic form factors using the Rosenbluth separation. 
We examine the independence of the extracted functions of the photon beam energy.
These functions, under additional assumptions, are related to the proton's gluon Gravitational Form Factors (gGFFs).
We compare the extracted functions with lattice calculations
of the gGFFs in the region of $0.5<|t|<2$~GeV$^{2}$, 
where they overlap.
Such analysis demonstrates the possibility of extracting some combinations of the 
gGFFs from the data at high $t$, complementary to the lattice calculations
available in the low $t$ region. 
However, higher statistics are needed to more accurately check the  predicted scaling 
behavior of the data and compare with the lattice results,
thus testing and comparing the theoretical assumptions used in 
the GPD and holographic models.
\end{abstract}

\maketitle

\section{Introduction}
Charmonium photoproduction near threshold provides a unique way to probe the gluonic structure 
of the nucleon.
Due to the high mass of the charm quark and proximity to the threshold, 
the reaction is expected to be dominated by two-gluon exchange \cite{Brodsky_Miller,Feng}.
In the GPD approach factorization is assumed
based on the high charmonium mass defining the hard scale in the reaction
\cite{Guo1,Guo2}.
As the two gluons can mimic graviton-like exchange, 
the gluonic GPD in turn can be related to the gluon Gravitational Form Factors (gGFFs) \cite{Guo1,Hatta_Strikman}.  
To be able to access the gluonic content of the nucleon,
the above chain of assumptions requires experimental testing. 

In the recent years, thanks to the 12 GeV Jefferson Lab electron accelerator, 
several experiments measured
the $J/\psi $ photoproduction near threshold with much higher precision
than the previous experiments from the 1970s.
The GlueX experiment in Hall D performed measurements with a nearly full acceptance spectrometer
in a wide range of photon beam energies starting from the threshold of $8.2$~GeV to the
maximum available energy of about $11.8$~GeV \cite{prl_gluex,prc_gluex}.
In addition to the total cross-section, the differential cross sections were extracted in three
slices of the photon beam energy, providing full coverage of the near-threshold kinematic region.
The $J/\psi$-007 experiment in Hall C accumulated similar statistics in a narrower
kinematic region with several settings of their two  low-acceptance spectrometers 
that however allowed very high luminosity \cite{hallc_jp007}. 
The differential cross sections were measured in 10 narrow energy bins with very high statistics 
in the forward region.

In Ref.~\cite{hallc_jp007} the gGFFs have been extracted by fitting the $J/\psi$-007 data with
the theoretical predictions from the holographic \cite{Zahed2} and GPD \cite{Guo1} approaches. 
In the GPD analysis of Ref.~\cite{Guo2}, the data from both 
GlueX and $J/\psi $-007 experiments were used to extract the gGFFs.
The GPD approach is based on expansion in the skewness 
parameter
$\xi$ for values approaching unity
(see also \cite{Hatta_Strikman}).
Therefore, its validity is justified for high $|t|$ values
due to kinematic correlation between $t$ and $\xi$.
A common feature of the above two analyses in Refs.~\cite{hallc_jp007,Guo2} is the  modeling of the gGFFs
with dipole/tripole functions. 
The dipole form was proposed in Ref.~\cite{strikman} 
in analogy with the electro-magnetic form factors,
whereas the tripole one is a good approximation of the holographic results \cite{Zahed2}.
Such functional forms are consistent with the
lattice calculations in Refs.~\cite{Shanahan1,Shanahan2} 
and used there as a parametrization of the results.
While the GPD approach works at high $|t|$, the lattice calculations of the gGFFs 
have been done so far in the $0<|t|<2$~GeV$^2$ region.
Therefore, the use of dipole/tripole parametrization,
especially at high $|t|$ requires experimental validation.
Most importantly, 
the extraction of the gGFFs in both analyses of Refs.~\cite{hallc_jp007,Guo2} 
is guided by additional external constraints
on some of the gGFF parameters, based on, or consistent with the lattice calculations.
As a result, first, the extracted gGFFs are sensitive 
to the normalization of the theoretical models and second,
when compared to the lattice calculations, 
such results are not completely independent. 

The analysis in this work is based on
one of the important results of Ref.~\cite{Guo3},
that the differential cross section of the charmonium photoproduction
for high $\xi $ values can be presented in this general form:
\begin{eqnarray}
d\sigma/dt &&  \sim \xi^{-4}\left[G_0(t)+\xi^2G_2(t)+\xi^4G_4(t)\right], 
\label{eq:G}
\end{eqnarray}
where the coefficients $G_0$, $G_2$, and $G_4$ are functions only of $t$, 
which in the leading-moment approximation are related to the gGFFs.
The holographic theory in Ref.~\cite{Zahed2} gives a similar expression: 
\begin{eqnarray}
d\sigma/dt &&  \sim \left[A_g(t)+\eta^24C_g(t)\right]^2, 
\label{eq:H}
\end{eqnarray}
making direct connection to the gravitational form factors $A_g(t)$
and $C_g(t)$,
where $\eta $ has a similar meaning as the skewness parameter $\xi $.
It is important to stress that the GPD and holographic theories have very different 
domains of validity.
The holographic theory is valid for soft processes because it works in the limit
of high coupling constant.
In contrast, the GPD theory requires high $|t|$ and works for hard processes.
We aim to extract the $t$-functions in Eqs.(\ref{eq:G},\ref{eq:H}) 
(we will call them form factor functions) from the experimental data
without any additional assumptions.
In contrast to the studies in Refs.~\cite{hallc_jp007, Guo2},
{\bf first, the procedure described in this work  does not use 
any external constraints and second,
the results will not be model functions fitted to the data,
but actual data points measuring the 
$t$-dependence of the above form factors.}

We point out the analogy with the Rosenbluth separation of the electric, $G_E$, and magnetic, $G_M$, form factors.
Based on the Rosenbluth formula \cite{Rosenbluth} 
for the electron-proton elastic cross section, $d\sigma/d\Omega$,
the method relies on the linear dependence of the reduced  cross section, $\sigma_R$,
on the kinematic variable $\epsilon $: 
\begin{equation}
\sigma_R = \frac{d\sigma}{d\Omega}/\left(\frac{d\sigma}{d\Omega}\right)_M \frac{\epsilon(1+\tau)}{\tau}=
\frac{\epsilon}{\tau}G_E^2(t)+G_M^2(t),
\label{eq:rosen}
\end{equation}
where $\left(\frac{d\sigma}{d\Omega}\right)_M$ is the Mott cross section, $\tau = -t/4m^2$, 
$\epsilon = 1/(1+2(1+\tau)tan^2\frac{\theta_e}{2})$,
where $m$ is the proton mass and $\theta_e$ is the electron scattering angle in the proton rest frame.
By measuring $\sigma_R$ at the same momentum transfer $t$ 
as function of $\epsilon /\tau$,
one can estimate the slope and the intercept, 
being respectively the electric and magnetic form factors squared
at this value of $t$.

The analogy of Eqs.(\ref{eq:G},\ref{eq:H}) with Eq.(\ref{eq:rosen}) is clear: instead of $\epsilon $,
we can use the variations with $\xi $ or $\eta $ at a fixed $t$ 
to extract the form factor functions in Eq.(\ref{eq:G},\ref{eq:H}).
In this paper we will apply a similar ``Rosenbluth'' technique,
however with some noticeable differences. 
Due to the limited experimental data so far, it is practically impossible to analyze
separately data sets in narrow ranges of $t$ versus some function of $\xi $ or $\eta $.
The GPD analysis has additional complications 
due to high $\xi $ requirement and the 
kinematic constraints that impose correlations between the ranges of $t$ and $\xi $.

In our analysis we will use the results from the two Jefferson Lab experiments, GlueX and $J/\psi$-007,
and in case of the GPD, selecting data points with $\xi >0.4$.
Such limit may not be high enough for the 
assumptions in Ref.~\cite{Guo3} to be valid,
and the selection is driven by the limited amount of data at high $\xi $.
In addition we will select measurements with photon energies above $9.3$~GeV,
to stay away from the open-charm thresholds of $\Lambda_c\bar{D}$ and $\Lambda_c\bar{D}^*$.
As discussed recently in Refs.\cite{openc,prc_gluex,JPAC_paper}, there is possible
evidence for contributions from reactions with open-charm exchange.

In section \ref{sec:rosen} we will formulate the ``Rosenbluth'' technique.
It will be used in section \ref{sec:CFF} to extract the form factor functions
and check their energy independence.
In section \ref{sec:GFF} we will work within the leading-term approximation to
demonstrate the possibility of extracting some combinations of the gGFFs 
and compare them to the lattice results.
Some additional remarks and summary will be given in section \ref{sec:sum}.

\section{The ``Rosenbluth'' technique}
\label{sec:rosen}
In the GPD calculations of Ref.~\cite{Guo3} the differential cross section of the $J/\psi $ photoproduction 
near threshold
is presented as a function of the skewness parameter $\xi $ and the momentum transfer $t$:
\begin{equation}
\frac{d\sigma}{dt}(\xi,t)
= F(E)\cdot \xi^{-4}\left[G_0(t)+\xi^2G_2(t)+\xi^4G_4(t)\right], 
\label{eq:general}
\end{equation}
where the factor $F$ includes a kinematic factor and the $J/\psi $ non-relativistic 
wave function at origin, 
and depends only on the photon beam energy, $E$,
while the $G$ functions depend only on $t$.
The skewness is defined as:
\begin{eqnarray}
\xi&&  = \frac{M^2-t}{2(s-m^2)-M^2+t}, 
\label{eq:xi}
\end{eqnarray}
where $M$ and $m$ correspond to the $J/\psi $ and the proton masses and
$s$ is the center of mass energy squared.
The $G$ functions are given by:
\begin{eqnarray}
G_0(t)&&=\left({\cal{A}}_g(t)\right)^2-\frac{t}{4m^2}\left({\cal{B}}_g(t)\right)^2 \label{eq:G0} \\
G_2(t)&&=2{\cal{A}}_g(t){\cal{C}}_g(t)+2\frac{t}{4m^2}{\cal{B}}_g(t){\cal{C}}_g(t) \nonumber \\
	&&-\left({\cal{A}}_g(t)+{\cal{B}}_g(t)\right)^2 \label{eq:G2} \\
G_4(t)&&=\left(1-\frac{t}{4m^2}\right)\left({\cal{C}}_g(t)\right)^2.
\label{eq:G4}
\end{eqnarray}
In the leading-moment approximation the functions
${\cal{A}}_g$, ${\cal{B}}_g$, and ${\cal{C}}_g$ are related to the 
gGFFs, 
$A_g$, $B_g$, and $C_g$ as:
\begin{eqnarray}
{\cal{A}}_g(t) &&= 2A_1^{conf}A_g(t) 
\label{eq:gCF0} \\
{\cal{B}}_g(t) &&= 2A_1^{conf}B_g(t)
\label{eq:gCF2} \\
{\cal{C}}_g(t) &&= 8A_1^{conf}C_g(t), 
\label{eq:gCF}
\end{eqnarray}
using the conformal expansion of the above functions, where $A_1^{conf}=5/4$ \cite{Guo3}.

The holographic approach of Ref.~\cite{Zahed2} gives direct relation between
the cross section and the gravitational form factors $A_g$ and $C_g$:
\begin{eqnarray}
	\frac{d\sigma}{dt}(\eta ,t) &&  = N(E)\left[A_g(t)+\eta^24C_g(t)\right]^2, 
\label{eq:ACH}
\end{eqnarray}
where
\begin{eqnarray}
\eta&&  = \frac{M^2}{2(s-m^2)-M^2+t}. 
\end{eqnarray}
In this approach the $B_g$ form factor is zero. 
The $N(E)$ function as given in Ref.~\cite{Zahed2} depends only on the beam 
energy\footnote{In the previous work of Ref.~\cite{Zahed4} this prefactor function
in the Eq.(8.6) has a weak dependence on $t$. 
As it varies no more than $10\%$ over the full kinematic region 
considered here, the $t$ dependence will be neglected.}
and is normalized using GlueX total cross section.  
This is in contrast to the GPD calculations that are absolute.
The holographic theory is valid in the double limit of strong gauge coupling
and large number of colors, $N_c$.
To indicate the approximate nature of the results 
we will rewrite the above formula in analogy with Eq.(\ref{eq:general}) as:
\begin{eqnarray}
	\frac{d\sigma}{dt}(\eta ,t) &&  = N(E)\left[H_0(t)+\eta^2H_2(t)+\eta^4H_4(t)\right], 
\label{eq:HH}
\end{eqnarray}
where in the limit of the validity of the theory we have:
\begin{eqnarray}
H_0(t) &&= A_g^2(t) 
\label{eq:gHFF0} \\
H_2(t) &&= 8A_g(t)C_g(t)
\label{eq:gHFF2} \\
H_4(t) &&= 16C_g^2(t), 
\label{eq:gHFF4}
\end{eqnarray}
In the GPD case we can rewrite Eq.(\ref{eq:general}) as:
\begin{eqnarray}
\frac{d\sigma}{dt}(\xi,t)
	&&= F(E)\cdot 4(A_1^{conf})^2\xi^{-4} \nonumber \\
	&&\times \left[G_0(t)+\xi^2G_2(t)+\xi^4G_4(t)\right]. 
\label{eq:gnorm}
\end{eqnarray}
Thus, in the leading-moment approximation and neglecting $B_g$, 
from Eqs.(\ref{eq:G0}-\ref{eq:gCF}) we obtain for the $G$ functions
the same normalization and
similar relations to the gGFFs as those for the $H$ functions:
\begin{eqnarray}
G_0(t) &&= A_g^2(t) 
\label{eq:gCFF0} \\
G_0(t)+G_2(t) &&= 8A_g(t)C_g(t)
\label{eq:gCFF2} \\
	G_4(t) &&= 16C_g^2(t)\left(1-\frac{t}{4m^2}\right). 
\label{eq:gCFF4}
\end{eqnarray}
In the further formulas we will attach the factor $4(A_1^{conf})^2$ 
to the $F(E)$ function.

We will derive the equations for the ``Rosenbluth'' separation in the GPD case only.
The formulas for the holographic approach can be derived in much the same way
and the results will be shown later.
From Eqs.(\ref{eq:gnorm}) we construct the reduced cross section:
\begin{eqnarray}
	\sigma_{R0}&&(E,t) = 
\frac{d\sigma}{dt}(E,t) \frac{\xi^2(E,t)}{F(E)} = \nonumber \\
	&&=  \xi^{-2}(E,t)G_0(t)+G_2(t)+\xi^2(E,t)G_4(t), 
\label{eq:GA2}
\end{eqnarray}
where we have indicated the energy and $t$ dependence of the corresponding 
variables\footnote{Note that all reduced cross sections defined in this work are dimensionless quantities}. 
We have verified, initially based on the lattice results from Refs.~\cite{Shanahan1,Shanahan2} 
discussed later,
that the last term in Eq.(\ref{eq:gnorm}), and respectively Eq.(\ref{eq:GA2}), 
is much smaller in absolute value than each of the other two terms.
Partially this is because of the $\xi^4$/$\xi^2$ suppression factor (as $\xi<1$)
with respect to the first/second term.
This assumption will be checked later based on the results of this analysis.
Note that this does not necessarily mean that the last terms
in Eqs.(\ref{eq:gnorm}) and (\ref{eq:GA2})
can be neglected.
It will be shown that the other two terms have different signs and it is possible 
they largely cancel each other.

The second term in Eq.(\ref{eq:GA2}) depends only on $t$ and we can eliminate it by taking
the difference of the cross sections at two different photon energies, $E_i$ and $E_j$:
\begin{eqnarray}
\sigma_{R0}(E_i,t) &&- \sigma_{R0}(E_j,t) 
 =  \left[\xi^{-2}(E_i,t)-\xi^{-2}(E_j,t)\right]G_0(t) \nonumber \\
&& + \left[\xi^{2}(E_i,t)-\xi^{2}(E_j,t)\right]G_4(t).
\label{eq:GA3}
\end{eqnarray}
The last term can be neglected with respect to the first one,
and we have the $G_0$ function extracted from experimental data:
\begin{eqnarray}
G_0^{exp.}(t) &&= 
	\left[\sigma_{R0}(E_i,t) - \sigma_{R0}(E_j,t)\right] \nonumber \\
&&/\left[\xi^{-2}(E_i,t)-\xi^{-2}(E_j,t)\right]
\label{eq:GA4}
\end{eqnarray}

We will choose a data set at a given energy $E_i$ as a reference.
Thus, the above equation will give us the slope 
of the reduced cross section as function of $\xi^{-2}$
for each additional data point with transfer momentum $t$ and 
energy $E_j$ with respect to the reference energy,
which is the $G_0(t)$ function and should not depend on the energy $E_j$.
Thus, checking the energy independence of the extracted form factor function
will serve as a test of the validity of the $\xi $-scaling as given by Eq.(\ref{eq:G}).
With this method
the uncertainties of the reference data set can be considered
to be systematic errors common for all data points.

The $G_2(t)$ function can be extracted in much the same way.
Here we give as an example the formulas for the sum $G_0(t)+G_2(t)$
that enters in Eq.~(\ref{eq:gCFF2}):
\begin{eqnarray}
(G_0(t)&&+G_2(t))^{exp.}  = \left[\sigma_{R02}(E_i,t) - \sigma_{R02}(E_j,t)\right] \nonumber \\
&&/\left[\frac{\xi^{2}(E_i,t)}{1-\xi^{2}(E_i,t)} - \frac{\xi^{2}(E_j,t)}{1-\xi^{2}(E_j,t)}\right] ,
\label{eq:G0G2}
\end{eqnarray}
where we have defined another reduced cross section as:
\begin{eqnarray}
\sigma_{R02}(E,t) = \frac{d\sigma}{dt}(E,t) \frac{\xi^4(E,t)}{F(E)(1-\xi^{2}(E,t))}. 
\label{eq:GR02}
\end{eqnarray}

\section{Energy independence of the form factor functions}
\label{sec:CFF}
\subsection{The GPD approach}
\begin{figure}[]
\includegraphics[width=0.45\textwidth]{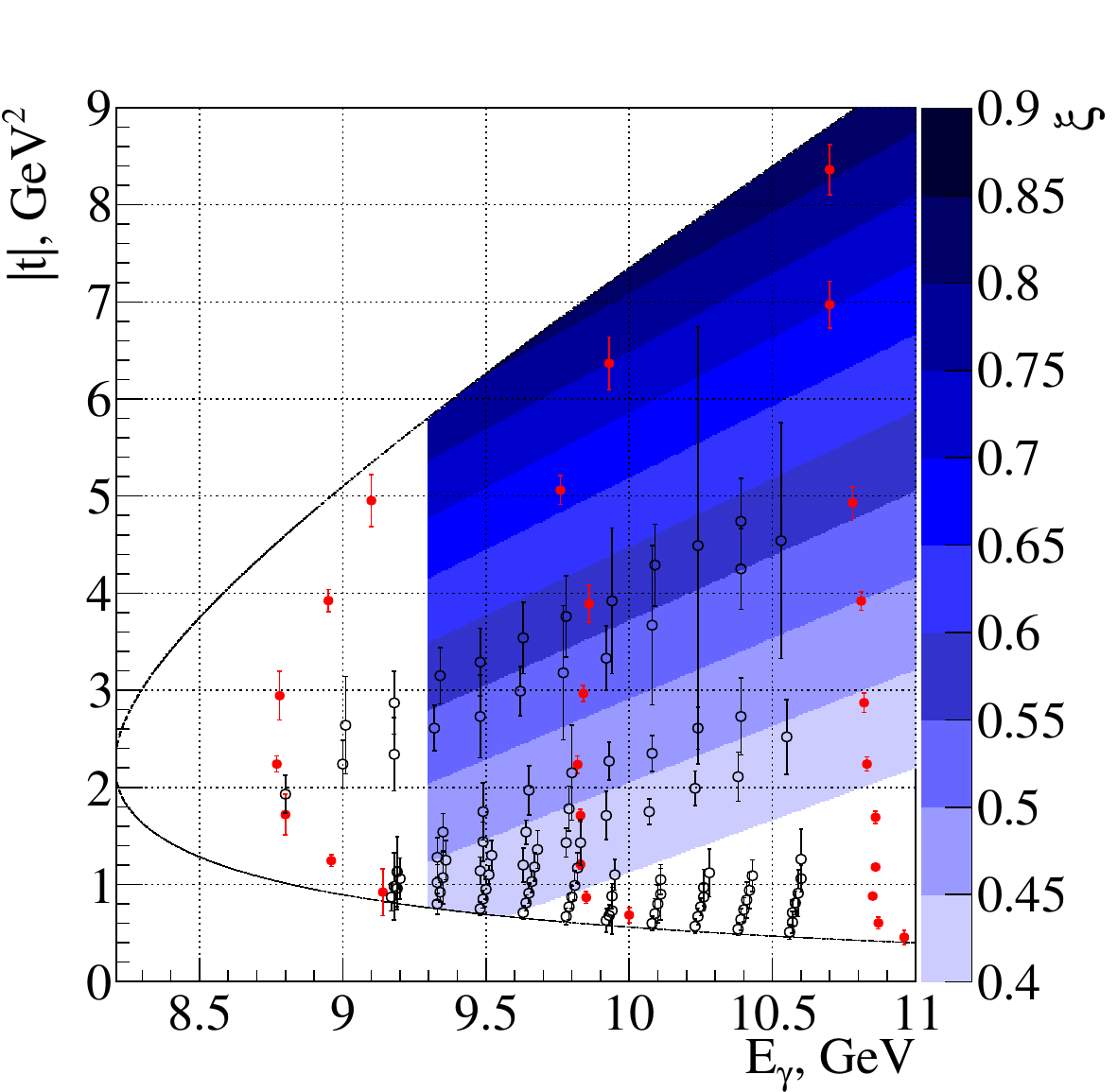}
  \caption{
Data points of the GlueX \cite{prc_gluex} (red solid) and $J/\psi$-007 \cite{hallc_jp007} (black open)
experiments on the $E$-vs-$t$ plane.
The vertical error bars represent the relative errors of the differential cross sections
measured at these points
in arbitrary units (not related to the $y$ axis).
The $\xi$ values on this plot are given on the $z$-axis (color coded).
Some of the $J/\psi$-007 points are slightly shifted up in energy for visibility.
The colored (shaded in print) area corresponds to the $\xi>0.4$ and $E>9.3$~GeV
region that is used in the GPD analysis. 
In the holographic case all the data points for $E>9.3$~GeV are used.
}
  \label{fig:2dkin}
\end{figure}
The practical implementation of the above technique is limited by the quality
and range of the available data.
The data sets from the two Jefferson Lab experiments are illustrated in Fig.\ref{fig:2dkin}.
As discussed above for the main part of the GPD analysis we will use the data with $\xi >0.4$
and $E>9.3$~GeV.
This region on the $E$-vs-$t$ plane in Fig.\ref{fig:2dkin} is shown with color.
We will choose the highest energy data set from the GlueX experiment, at 
an average energy $E_i=10.82$~GeV,
as a reference and subtract the other data at different energies $E_j$.
Such a choice gives maximum leverage in $\xi $ when calculating the slopes 
from Eqs.(\ref{eq:GA4}) and (\ref{eq:G0G2}), at the same time the 
highest energy data set is more precise and has the widest range in $t$.
\begin{figure}[]
\includegraphics[width=0.45\textwidth]{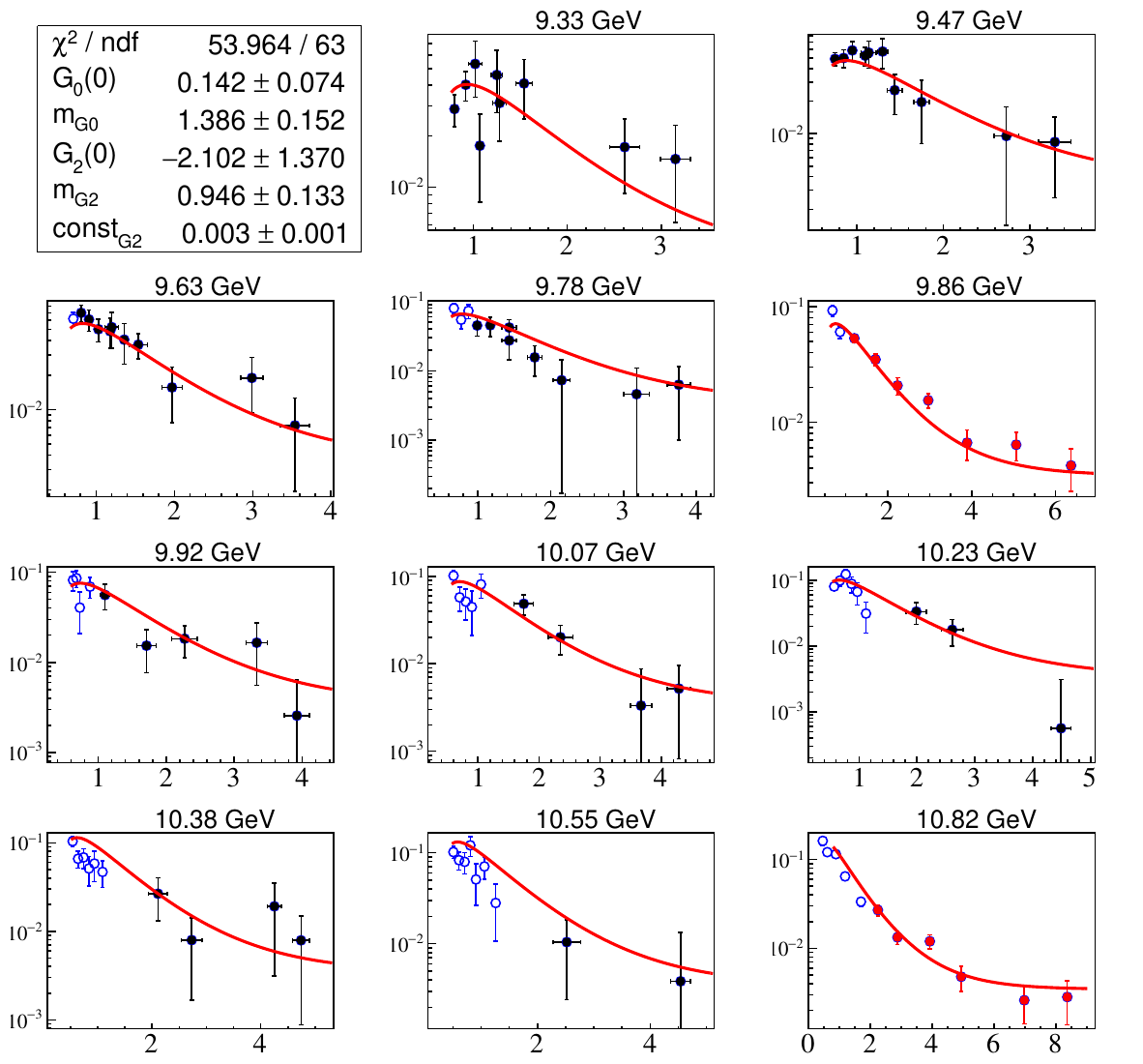}
  \caption{
Global five-parameter fit of the reduced cross sections, $\xi^2/F \cdot d\sigma/dt$, 
	as function of $|t|$ (in GeV$^2$), using the GPD Eq.(\ref{eq:gnorm})
and the parametrizations given by
Eqs.~(\ref{eq:pG0},\ref{eq:pG2}),
from GlueX (solid red at $9.86$ and $10.82$~GeV) and $J/\psi$-007 (solid black) experiments
for different photon energies as indicated.
The data points with $\xi <0.4$ are plotted (open blue) but not used in the fit.
For the further analysis only the fitted function at $10.82$~GeV is used.
}
  \label{fig:fits}
\end{figure}

In principle, we can fit only the reference cross section with some function of $t$,
so that we can use it in Eqs.(\ref{eq:GA4}) and (\ref{eq:G0G2})
to subtract the reduced cross sections
at the $t$ values of the data points at the other energies.
Still, as seen in Fig.\ref{fig:2dkin}, due to the $\xi >0.4$ constraint, the $1-2$~GeV$^2$ region in $|t|$ 
is not covered by the reference cross section, therefore it requires extrapolation to lower $|t|$.
For that we will fit all the cross sections simultaneously using Eq.~(\ref{eq:gnorm}).
Thus, the extrapolation of the reference energy will be guided by the data at
the other energies, at the same time reducing the error of the function
fitted to the reference energy.

We demonstrate this procedure when fitting the reduced
cross section, $\sigma_{R0}$, as function of $t$ for all energies
using Eq. (\ref{eq:gnorm}). 
Good fit results are obtained 
if we neglect $G_4$ and parametrize 
$G_0$ and $G_2$ with squared dipole functions:
\begin{eqnarray}
	G_0(t) &&= \frac{G_0(0)}{(1-t/m_{G0}^2)^4}
\label{eq:pG0} \\
	G_2(t) &&= \frac{G_2(0)}{(1-t/m_{G2}^2)^4}+conts_{G2}.
\label{eq:pG2} 
\end{eqnarray}
The constant term added for $G_2$ 
improves the fit quality for $|t|>5$~GeV$^2$
and it is not important for the extrapolation 
of the reference cross section to lower $|t|$ values.
The fit results are shown in Fig.\ref{fig:fits}.
The data points for $\xi<0.4$, also plotted in the figure but not used
in the fit, deviate from the fitted functions.
In the further analysis for the extraction of $G_0^{exp.}(t)$, in Eq.(\ref{eq:GA4})
we will use only 
the fitted function for the reference energy 
$E_i=10.82$~GeV (Fig.\ref{fig:fits} bottom right).

\begin{figure}[]
\includegraphics[width=0.45\textwidth]{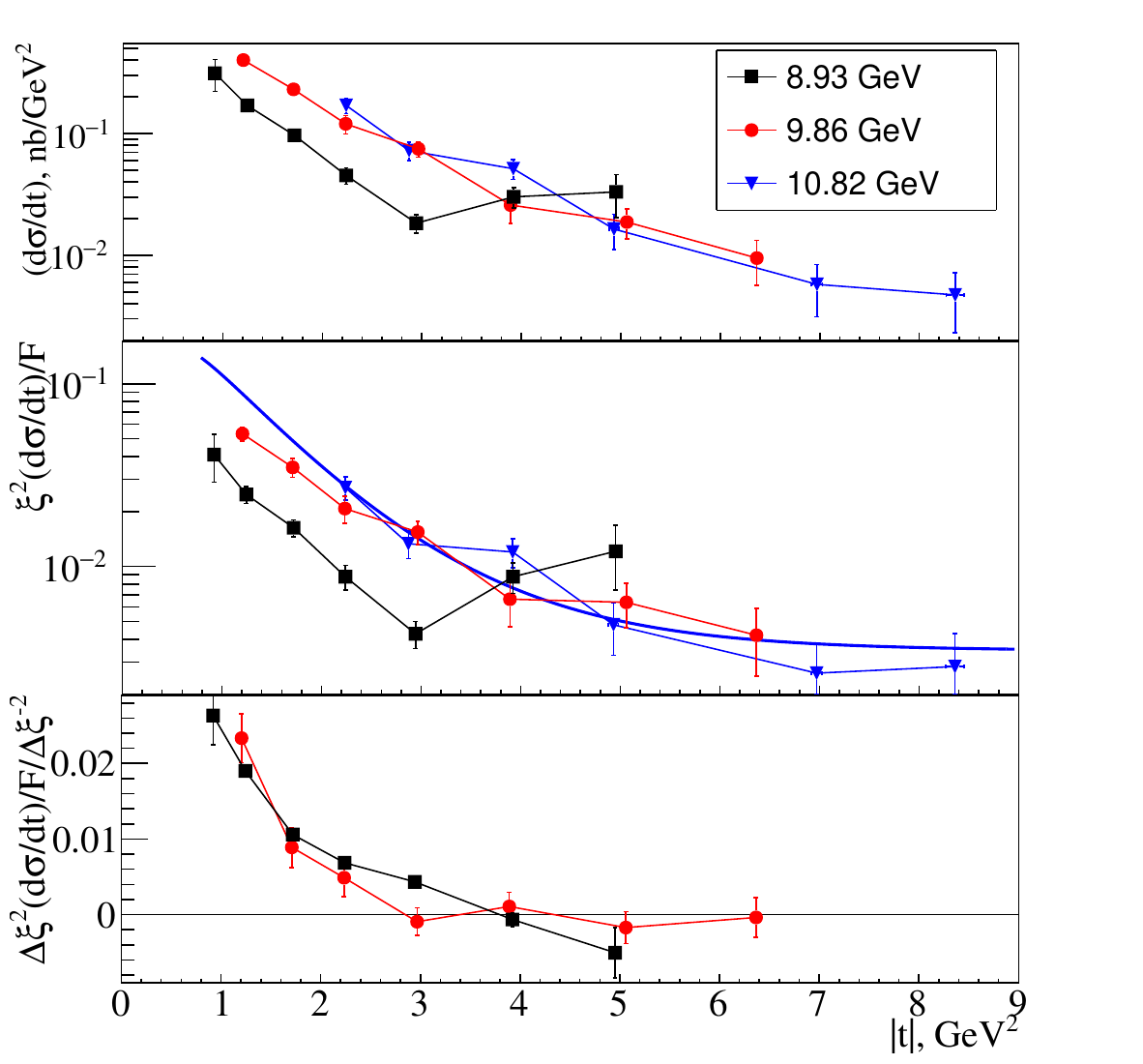}
  \caption{
Different stages of applying the ``Rosenbluth'' extraction of the $G_0(t)$ function
using the GlueX data:
the differential cross sections at three energy slices (top),
the reduced cross sections - Eq.(\ref{eq:GA2}) (middle),
and the extracted $G_0(t)$ function with Eq.(\ref{eq:GA4}) 
using the highest energy as a reference (bottom).
The fitted function of the reduced cross section at the highest energy 
from the global fit (Fig.\ref{fig:fits} bottom right),
is shown in the middle panel (blue curve), as well.
The points are connected with lines to guide the eye only.
}
  \label{fig:3comp}
\end{figure}

\begin{figure}[h]
 \begin{subfigure}[b]{0.45\textwidth}
  \includegraphics[width=\textwidth]{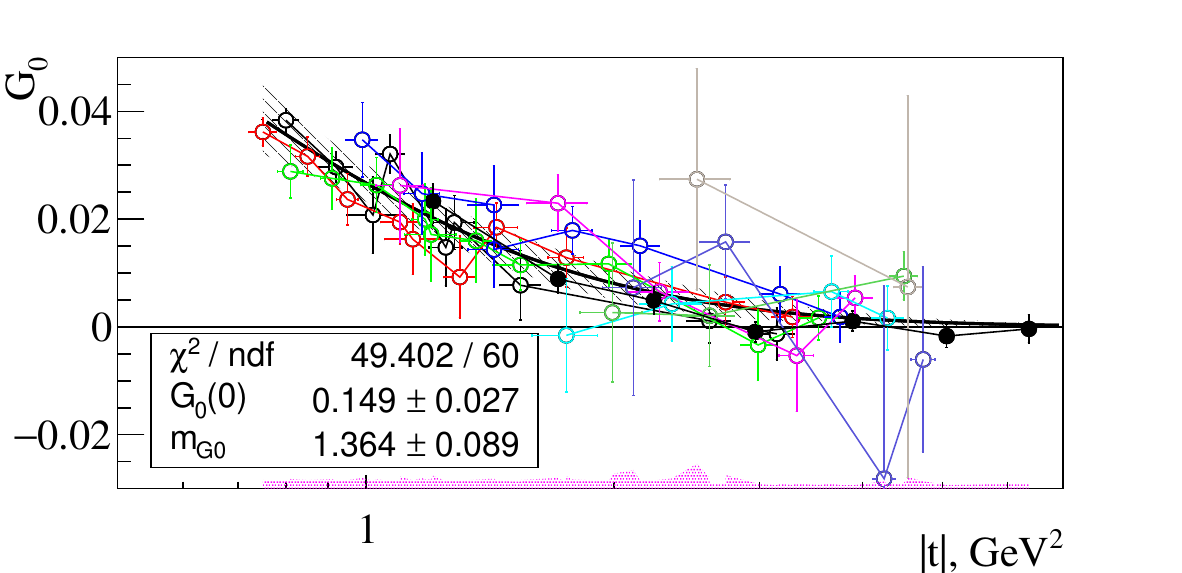}
  \caption{}
  \label{fig:G0_E}
  \end{subfigure}
 \begin{subfigure}[b]{0.45\textwidth}
\includegraphics[width=\textwidth]{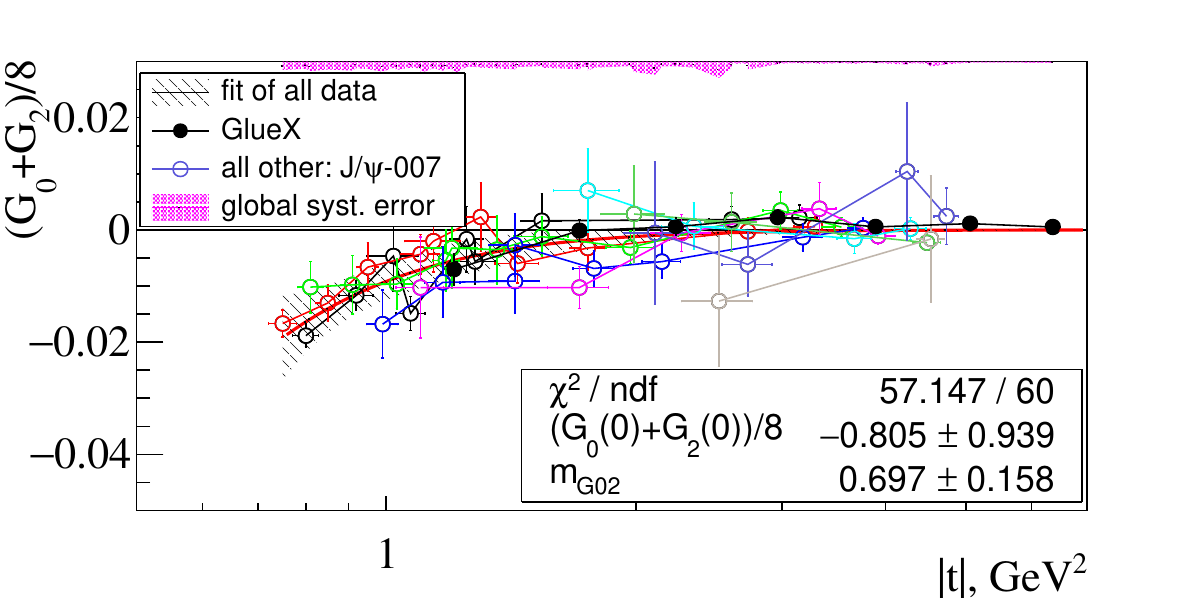}
  \caption{}
  \label{fig:G2_E}
  \end{subfigure}
  \caption{
	  $G_0(t)$ (top) and $(G_0(t)+G_2(t))/8$ (bottom) functions as extracted from the GlueX (solid)
and $J/\psi$-007 (open symbols) data for $\xi>0.4$ using Eqs.(\ref{eq:GA4},\ref{eq:G0G2})
for different photon beam energies (different colors connected with lines). 
The data points are fitted with squared dipole functions, 
$G(0)/(1-t/m_G^2)^4$, 
shown with confidence range at $95\%$ confidence level.
}
  \label{fig:G02}
\end{figure}

To illustrate clearly how the ``Rosenbluth'' technique works,
in Fig.~\ref{fig:3comp} we show the different stages of the 
extraction of the $G_0(t)$ function from the GlueX data.
Here, just for this demonstration, we have included the differential cross section at
the lowest beam energy. 
Starting from the original cross sections, 
going to the $\xi$-scaled cross sections, and finally
calculating the slopes from  Eq.(\ref{eq:GA4}), 
it is remarkable to see how the results at the different energies 
converge into a function, $G_0^{exp.}(t)$, that is largely energy independent.

The results for $G_0$ and the sum $G_0+G_2$ using all the data  
are shown in Figs.~\ref{fig:G0_E} and \ref{fig:G2_E}.
For better comparison of the two results, we have divided the sum by $8$ 
- see Eqs.(\ref{eq:gCFF0},\ref{eq:gCFF2}).
We see that the extracted functions do not depend on energy within the experimental errors.
Numerically, this is confirmed by the good quality of the fits of all the data using
squared dipole functions.
The uncertainties of the fitted function at the reference energy 
(as in Fig.~\ref{fig:fits} bottom right panel)
are scaled by the factors 
in Eqs.(\ref{eq:GA4},\ref{eq:G0G2}) and shown in Figs.\ref{fig:G0_E} and \ref{fig:G2_E}
as global systematic errors.

\subsection{Holographic approach}
\begin{figure}[]
\includegraphics[width=0.45\textwidth]{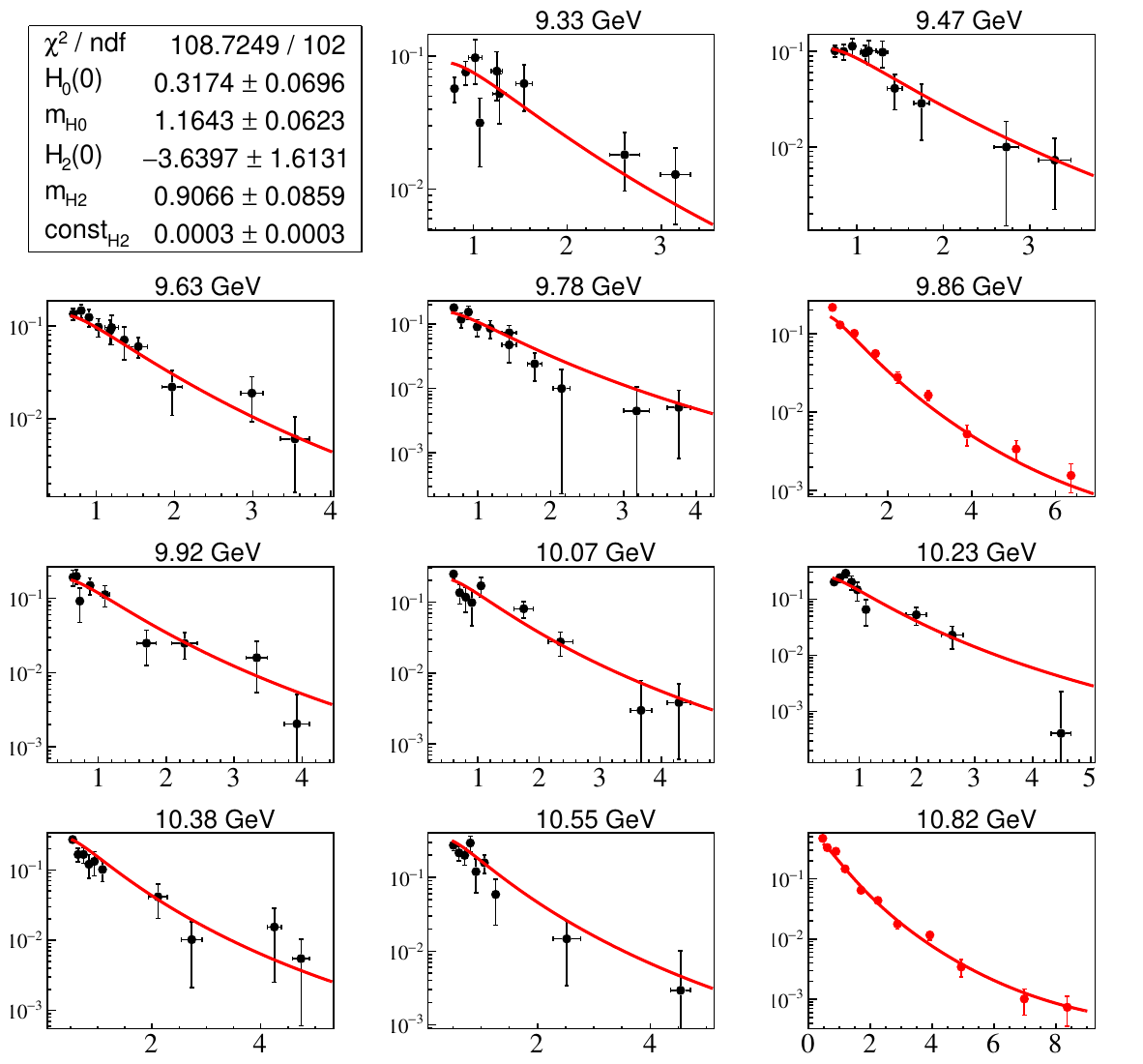}
  \caption{
	  Global five-parameter fit of the reduced cross sections, $\eta^{-2}/N \cdot d\sigma/dt$, 
	as function of $|t|$ (in GeV$^2$), using the holographic 
	Eq.(\ref{eq:HH}),
from GlueX (solid red at $9.86$ and $10.82$~Ge) and $J/\psi$-007 (solid black) experiments
for different photon energies as indicated.
For the further analysis only the fitted function at $10.82$~GeV is used.
}
  \label{fig:hfits}
\end{figure}
The formulas for extracting the $H_0$ and $H_2$ form factor functions
are the same as for $G_0$ and $G_2$ when replacing $\xi $ by $\eta $, except for the definition of the
reduced cross sections. 
The $H_0$ function is extracted as: 
\begin{eqnarray}
	H_0^{exp.}(t) &&= 
\left[\sigma_{H0}(E_i,t) - \sigma_{H0}(E_j,t)\right] \nonumber \\
&&/\left[\eta^{-2}(E_i,t)-\eta^{-2}(E_j,t)\right],
\label{eq:HA0}
\end{eqnarray}
where the corresponding reduced cross section is defined as:
\begin{eqnarray}
\sigma_{H0}(E,t) = 
	\frac{d\sigma}{dt}(E,t) \frac{\eta^{-2}(E,t)}{N(E)}.
\label{eq:RH0}
\end{eqnarray}
When compared to the GPD calculations, the difference comes from the extra $\xi^{-4}$ prefactor
in Eq.~(\ref{eq:general}).

Following the same procedure as in the case of the GPD analysis,
we perform global fit of the reduced cross sections using Eq.~(\ref{eq:HH}),
where $H_0$ and $H_2$ functions are parametrized in the same way as
$G_0$ and $G_2$ in Eqs.~(\ref{eq:pG0},\ref{eq:pG2}) and $H_4$ is neglected.
The most important difference is that in the holographic case all the data
points with $E >9.3$~GeV are included in the fit and there is no
need to extrapolate the reference cross section at $E_i=10.82$~GeV
as its $t$-range covers
the ranges of all the other measurements, see Fig.~\ref{fig:2dkin}.
The fit results shown in Fig.~\ref{fig:hfits} demonstrate good fit quality.

\begin{figure}[]
 \begin{subfigure}[b]{0.45\textwidth}
\includegraphics[width=\textwidth]{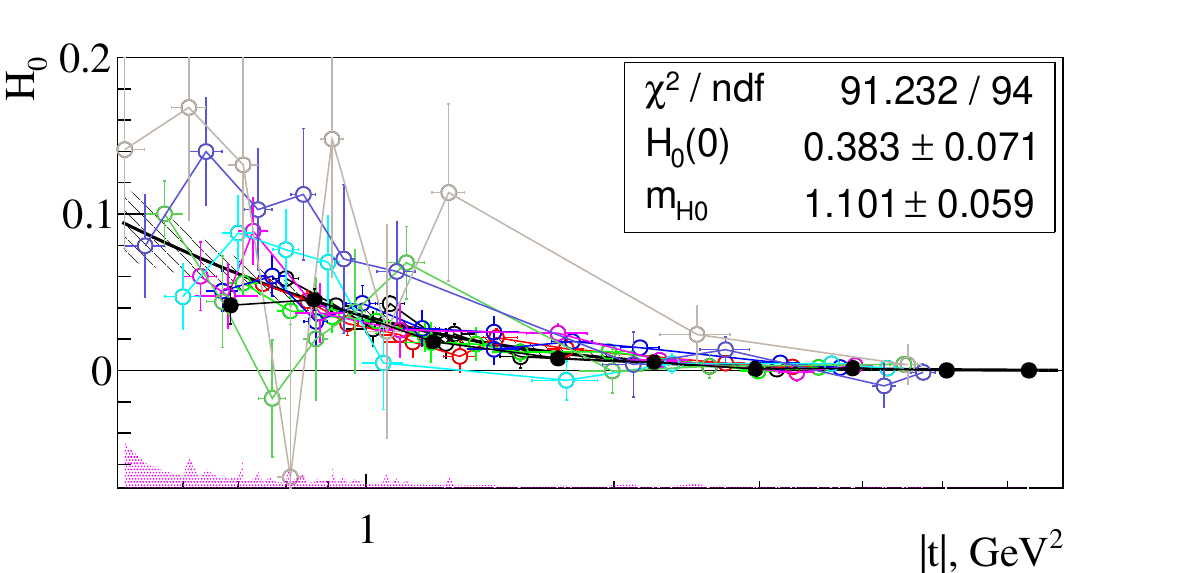}
  \caption{}
  \label{fig:G0_xi}
  \end{subfigure}
  \begin{subfigure}[b]{0.45\textwidth}
\includegraphics[width=\textwidth]{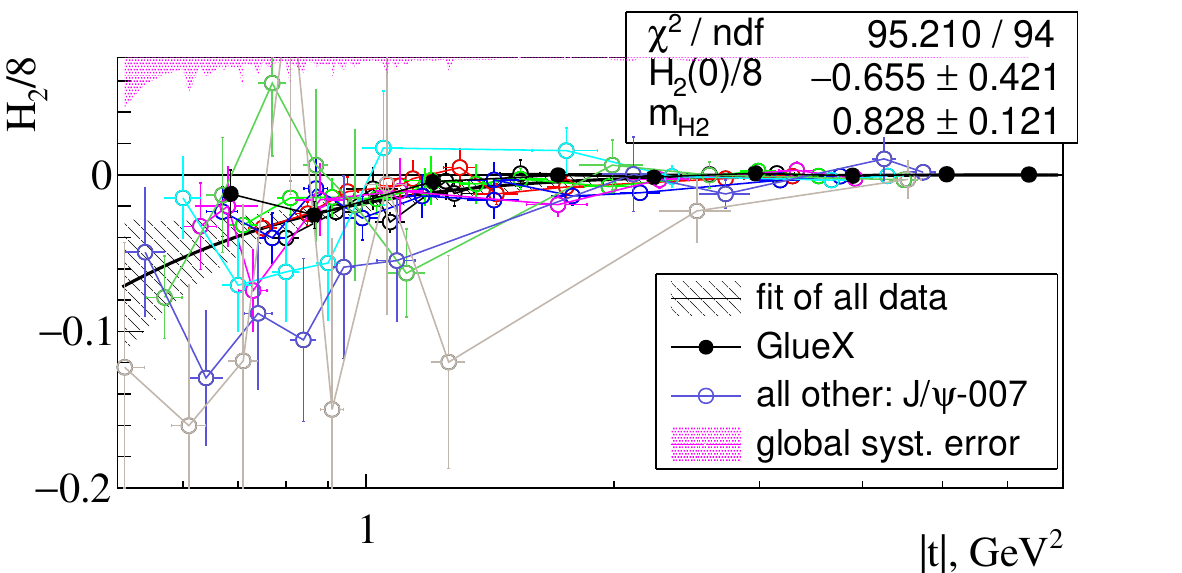}
  \caption{}
  \label{fig:G2_xi}
  \end{subfigure}
  \caption{
$H_0(t)$ (top) and $H_2(t)/8$ (bottom) functions as extracted from the GlueX (solid)
and $J/\psi$-007 (open symbols) data using Eqs.(\ref{eq:HA0},\ref{eq:HA2})
for different photon beam energies (different colors connected with lines). 
The data points are fitted with squared dipole functions, 
$H(0)/(1-t/m_H^2)^4$,
shown with confidence range at $95\%$ confidence level.
}
	\label{fig:H02}
\end{figure}
We extract the $H_2$ function in a very similar way:
\begin{eqnarray}
H_2^{exp.}(t) &&= 
\left[\sigma_{H2}(E_i,t) - \sigma_{H2}(E_j,t)\right] \nonumber \\
&&/\left[\eta^{2}(E_i,t)-\eta^{2}(E_j,t)\right],
\label{eq:HA2}
\end{eqnarray}
where the corresponding reduced cross section is just the normalized 
differential cross section:
\begin{eqnarray}
\sigma_{H2}(E,t) = 
	\frac{d\sigma}{dt}(E,t) \frac{1}{N(E)}.
\label{eq:RH2}
\end{eqnarray}
The results for the extracted $H_0$ and $H_2$ form factor functions 
are shown in Fig.~\ref{fig:H02}. 
They are fitted with squared dipole functions, 
demonstrating that in the case of the holographic
calculations, the extracted data points are also energy independent. 
As there is no constraint on the skewness parameter,
the extracted form factor functions extend down to lower $|t|$ values,
as compared to the GPD case.
Note, however, that especially in the low $|t|$ region the errors are much bigger
than in the GPD analysis, as the errors of the measurements
are scaled by bigger factors in Eqs.(\ref{eq:RH0},\ref{eq:RH2}),
compared to Eqs.(\ref{eq:GA2},\ref{eq:GR02}),
in the corresponding reduced cross sections.

\section{Relation to the gluon Gravitational Form Factors}
\label{sec:GFF}
In the leading-moment approximation the $G$ functions are related 
to the gGFFs according to Eqs.(\ref{eq:gCFF0}-\ref{eq:gCFF4}),
where we have neglected the $B_g$ form factor.
In case of the holographic theory, if the required conditions for the theory
to be valid are met, the $H$ functions are related to the gGFFs
as given by Eqs.(\ref{eq:gHFF0}-\ref{eq:gHFF4}).

\begin{figure}[]
 \begin{subfigure}[b]{0.45\textwidth}
\includegraphics[width=\textwidth]{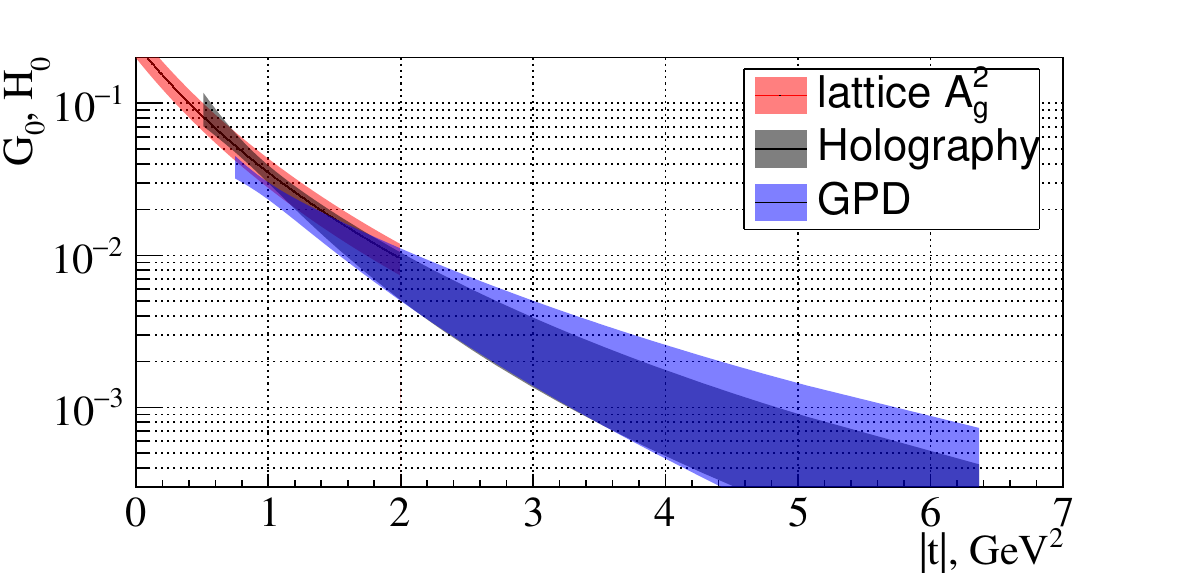}
  \caption{}
\label{fig:lattice_a2}
  \end{subfigure}
 \begin{subfigure}[b]{0.45\textwidth}
\includegraphics[width=\textwidth]{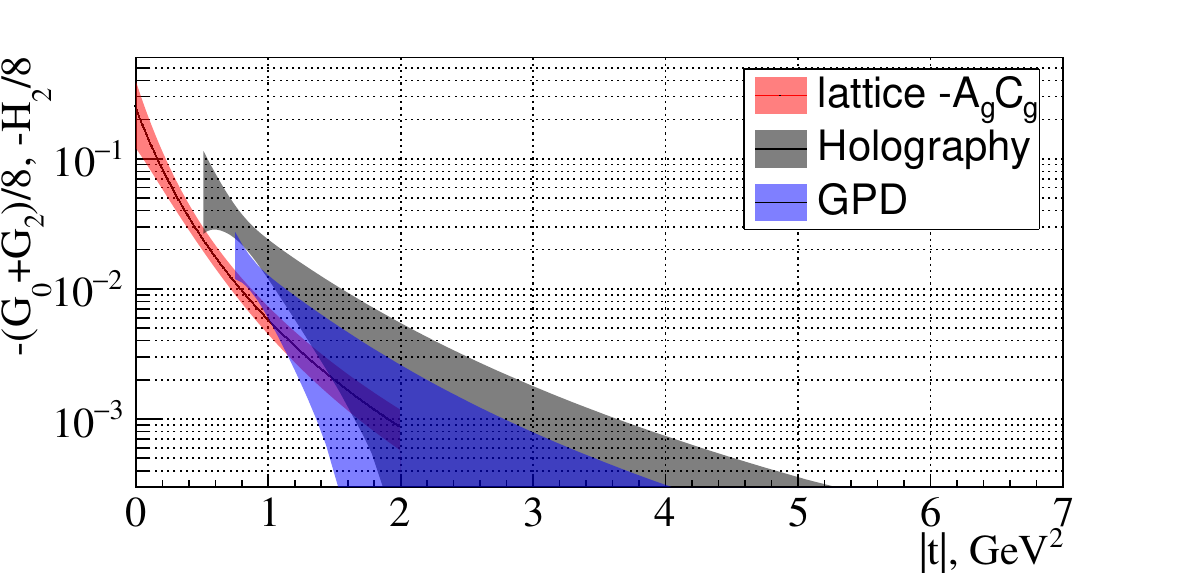}
  \caption{}
\label{fig:lattice_ac}
  \end{subfigure}
  \caption{
Comparison of the functions fitted to the extracted data points 
from Figs.~\ref{fig:G02},\ref{fig:H02}
with the corresponding lattice calculations of Ref.~\cite{Shanahan2}:
(top) $G_0$, $H_0$, compared to $A_g^2$ from lattice,
(bottom) $-(G_0+G_2)/8$, $-H_0/8$, compared to $-A_gC_g$ from lattice.
All the uncertainty bands correspond to $95\%$ confidence level.
}
\label{fig:lattice}
\end{figure}

In Fig.~\ref{fig:lattice_a2} we compare the functions fitted 
to the data points (as extracted in the previous section) for $G_0$ and $H_0$.
We see a general agreement between the GPD and holographic results,
which is remarkable as these two theories are diametrically different.
A possible way to interpret their agreement is that the corrections
that have very different nature for the two theories, are not dominant.
Indeed, according to Eqs.(\ref{eq:gCFF0},\ref{eq:gHFF0}),
the leading terms of both, $G_0$ and $H_0$, are the same and equal to 
the $A_g$ form factor squared. 
This assumption is supported by comparing
the experimental results to
the lattice calculations of $A_g^2$
from Ref.~\cite{Shanahan2} (shown in the same figure), where we
see an agreement at least within $2\sigma $ in the common $0.5-2$~GeV$^2$ region.

The leading terms of $G_0+G_2$ and $H_2$
should be proportional to the product 
of the gGFFs $A_gC_g$ (Eqs.(\ref{eq:gCFF2},\ref{eq:gHFF2})).
The functions fitted to data points for $(G_0+G_2)/8$ and $H_2/8$
from the previous section are shown in Fig.~\ref{fig:lattice_ac}
and compared to the lattice calculation of $A_gC_g$.
The agreement is not as good as in the $A_g^2$ case,
at the same time the uncertainties of the data are much bigger so that
we cannot make a definitive conclusion.

We will discuss the vanishing of the terms containing $G_4$ and $H_4$
in Eqs.(\ref{eq:gnorm},\ref{eq:HH})
within the leading-term approximation used in this section.
Note that the mass scale parameter for $G_0$ is bigger than the one for $G_0+G_2$, $m_{G0}>m_{G02}$ (Fig.~\ref{fig:G02}).
The same is true when comparing the results for $H_0$ and $H_2$: $m_{H0}>m_{H2}$ (Fig.~\ref{fig:H02}).
According to Eqs.(\ref{eq:gCFF0},\ref{eq:gCFF2}) and Eqs.(\ref{eq:gHFF0},\ref{eq:gHFF2}),
this means that $C_g$ vanishes faster with $|t|$ than $A_g$. 
As $G_4$ and $H_4$ are proportional to $C_g^2$ (Eqs.(\ref{eq:gCFF4},\ref{eq:gHFF4})),
this explains qualitatively their vanishing contribution 
in Eqs.(\ref{eq:gnorm},\ref{eq:HH}) at high $|t|$ values.
At low $|t|$ where $\xi$ and $\eta$ are much smaller than unity,
these terms are suppressed by the $\xi^{4}$/$\xi^{2}$,
or $\eta^{4}$/$\eta^{2}$ factors when compared to the first/second term
in Eqs.(\ref{eq:gnorm},\ref{eq:HH}).
From the extracted $A_g^2$ and $A_gC_g$ combinations we can evaluate separately the
$C_g$ form factor and then calculate the $G_4$ and $H_4$ functions 
using Eqs.(\ref{eq:gCFF4},\ref{eq:gHFF4}).
In this way we have verified quantitatively, based on the results in this paper, that for $|t|>0.7$~GeV$^2$,
$\xi^{4}G_4$ is about an order of magnitude smaller in absolute value than $G_0$ and $\xi^{2}G_2$.
The same is true in the holographic case:
$\eta^{4}H_4$ is much smaller in absolute value than $H_0$ and $\eta^{2}H_2$
for $|t|>0.5$~GeV$^2$.
Thus, in the leading-term approximation, this justifies neglecting the $G_4$/$H_4$
terms when extracting the form factors.

\section{Additional remarks and summary}
\label{sec:sum}
An interesting result can be deduced from the general $\xi$-scaling expression, Eq.(\ref{eq:general}),
though not directly related to the extraction of the form factors.
First, we note that $G_0$ must be positive as a sum of squares of ${\cal{A}}_g$ 
and ${\cal{B}}_g$ functions, according to Eq.(\ref{eq:G0})
with $t$ being negative. 
Therefore, from Eq.(\ref{eq:GA4}), for $E_i>E_j$ 
at a fixed $t$, as $\xi(E_i,t)<\xi(E_j,t)$ and therefore $\xi^{-2}(E_i,t)-\xi^{-2}(E_j,t)>0$, 
it follows that: 
\begin{eqnarray}
	d\sigma/dt(E_i,t)\frac{\xi^2(E_i,t)}{F(E_i)}&&>d\sigma/dt(E_j,t)\frac{\xi^2(E_j,t)}{F(E_j)}, \\
	&& E_i>E_j. \nonumber
\label{eq:dsdt_Er}
\end{eqnarray}
In particular, as $\xi^2(E_i,t)/F(E_i)<\xi^2(E_j,t)/F(E_j), $\footnote{
	This inequality follows from Eq.(\ref{eq:xi}) and the kinematical factor 
	$F(E)$ being proportional to $1/(s-m^2)^2$ \cite{Guo3}.
}
it means that:
\begin{eqnarray}
d\sigma/dt(E_i,t)>d\sigma/dt(E_j,t),\; E_i>E_j.
\label{eq:dsdt_E}
\end{eqnarray}
In other words, the differential cross section at a fixed $t$ must increase with the energy,
something that is not trivial. 

In the GPD approach our ``Rosenbluth'' procedure relies on Eq.(\ref{eq:general}) that comes in Ref.~\cite{Guo3} 
from some general theoretical assumptions about the $J/\psi$ photoproduction mechanisms,
such as gluon exchange and factorization.  
The observation of the energy independence of the extracted $G$ form factor functions
means that the experimental data, within their uncertainties,
are consistent with the assumptions in Ref.~\cite{Guo3}.
In the holographic case, 
Eq.(\ref{eq:H}), which relates directly $d\sigma /dt$
to the gGFFs can be interpreted as a more general  
expression in powers of $\eta $ as in Eq.(\ref{eq:HH}).
The energy independence of the extracted $H$ functions supports such an interpretation
within the errors of the present data set.
We stress that these results are based only on Eqs.(\ref{eq:general},\ref{eq:HH}) 
and no additional information is used. 
Note that any constraints on the gGFFs like fixing parameters on their parametrizations
depend on the assumption of a leading-term approximation.

The results from the ``Rosenbluth'' separation 
for the parameters of the functions fitted to $G_0$, $H_0$, and $H_2$ data points
(Figs.~\ref{fig:G02},\ref{fig:H02})
and the results of the global fits in Figs.~\ref{fig:fits},\ref{fig:hfits}
are consistent with each other\footnote{
The ``Rosenbluth'' results for $G_2$ are also consistent with 
the fit parameters in Fig.~\ref{fig:fits}. 
For the sake of focusing on the analogy between the GPD and holographic approaches,
these results are not shown in the paper, 
as it is the sum $G_0+G_2$ that has the same relation to the gGFFs as the $H_2$ function.}.
This is a result of using the same data set and the same parametrization for
the $G$ and $H$ functions.
However, these two methods of extracting the $G$ and $H$ functions are not equivalent.
With the ``Rosenbluth'' separation we extract data points, while the 
global fit method assumes a priori the functional forms 
used to parametrize the $G$ and $H$ functions.
The ``Rosenbluth'' separation uses only the function fitted to the reference
data set. 
The fits to all the data used in this case
are needed to reduce the systematic errors 
and, in the GPD case, to extrapolate 
the reference data to lower $|t|$ values.
In the holographic case we can extract the $H$ functions by interpolating 
the reference cross section without fitting all the data.
The results extracted in this way are similar to those presented above, except for the
bigger systematic uncertainties.

We conclude, based on the analysis of the current experimental data (Fig.\ref{fig:G02}),
that the $\xi$-scaling behavior is valid for $\xi >0.4$.
As discussed in Ref.~\cite{Guo3}, such a low limit is not justified theoretically 
and may result in a significant bias.
However, this value should be considered as a lower limit of the validity of the $\xi$-scaling,
as more precise data could move this value up.

In Fig.~\ref{fig:lattice} we have compared the functions fitted to the $G$ and $H$ data points
with  the calculations in the leading-term approximation
when using lattice gGFFs.
The procedure described in this paper does not use 
any external constraints.
The absolute value of the extracted form factor functions comes 
from the slope of the reduced cross sections
as a function of the energy.
Thus, we find it remarkable that the data points 
obtained with two diametric theories
are on the same scale as the lattice results.
If with higher statistics 
the agreement between the data and lattice calculations in the overlapping $t$-region is confirmed,
this could justify the extraction of the gGFFs at high $t$ directly from the data,
complementary to the lattice results available at low $t$. 

In a recent paper the GPD framework is extended by including Next-to-Leading-Order (NLO) calculations \cite{Guo4}.
The authors infer the GFFs for gluons and quarks (the latter enter in the  NLO diagrams),
parametrized with dipole/tripole functions, 
by fitting the data and lattice results separately
and simultaneously.
When fitting the data only, constraints on the gluon/quark GFFs $A_{g,q}(0)$ are applied and the results 
are found to be generally compatible with the lattice fits.
The Jefferson Lab data for $\xi >0.5$ are used in their analysis.
The theoretical error at this limit of the skewness parameter is estimated to be $\approx 30\%$.
The work presented in our paper can be considered to be an independent and complementary approach 
that confirms the
agreement between the data and lattice results within the present 
theoretical and experimental uncertainties.

It is critical to further investigate the near-threshold region of $E<9.3$~GeV.
For that we would need more statistics and further theoretical and phenomenological studies
to understand the possible contribution of open charm channels. 
If gluon exchange still dominates at least in some kinematic region,
it will be very important to have this region included in the presented analysis.
It is believed that close to threshold, being close to both $|t|_{min}$ and $|t|_{max}$
kinematic limits, the factorization would be better justified, at the same time the 
higher $\xi$ values in this region would better satisfy the requirements for the high-$\xi$
expansion in Ref.\cite{Guo3}.

The results in this paper should not be understood
as a  proof of the validity of the assumptions in Ref.~\cite{Guo3} and Ref.~\cite{Zahed2},
rather just as an indication, within the present data uncertainties, 
of the validity of the scaling as defined by  Eqs.(\ref{eq:general},\ref{eq:HH}).
Such scaling behavior as demonstrated experimentally in this work 
can be considered as an independent phenomenological observation
possibly coming from some general expansion in the skewness parameter.
In any case, the presented results are encouraging and
should stimulate the continuation of the experimental studies and further theoretical analysis.

\section{ACKNOWLEDGMENTS}
We would like to thank Yuxun Guo, Kiminad Mamo, and Ismail Zahed for the fruitful discussions
and explanations of the theories on which the results of this work are based
and Simon Taylor for valuable comments.
We thank Dimitra Pefkou for providing us with parametrizations of the lattice results.

Notice: Authored by Jefferson Science Associates, LL under U.S. DOE Contract
No. DE-AC05-06OR23177. The U.S. Government retains a non-exclusive, paid-up, Eire-
vocable, world-wide license to publish or reproduce this manuscript for U.S. Government
purposes.

\bibliography{jpsi_GFF.bbl}

\end{document}